\documentclass{article}

\PassOptionsToPackage{numbers, compress}{natbib}

\usepackage[final]{neurips_2021}

\usepackage[utf8]{inputenc} % allow utf-8 input
\usepackage[T1]{fontenc}    % use 8-bit T1 fonts
\usepackage{hyperref}       % hyperlinks
\usepackage{url}            % simple URL typesetting
\usepackage{booktabs}       % professional-quality tables
\usepackage{amsfonts}       % blackboard math symbols
\usepackage{nicefrac}       % compact symbols for 1/2, etc.
\usepackage{microtype}      % microtypography
\usepackage{xcolor}         % colors
\usepackage{microtype}
\usepackage{graphicx}
\usepackage{subfigure}
\usepackage{amsmath}
\usepackage{CJK}

\newcommand{\NAME}{\texttt{zephyr} }

\title{\texttt{zephyr} : Stitching Heterogeneous Training Data with Normalizing Flows for Photometric Redshift Inference}

\author{%
  Zechang Sun (孙泽昌)\thanks{\href{https://zechangsun.github.io}{https://zechangsun.github.io}} \\
  Department of Astronomy\\
  Tsinghua University \\
  Beijing, China 100084 \\
  \texttt{szc22@mails.tsinghua.edu.cn} \\
  \And
  Joshua S. Speagle (沈佳士)\thanks{Sencodary Affiliations: Department of Astronomy \& Astrophysics, University of Toronto, Toronto, ON, Canada; Data Sciences Institute, University of Toronto, Toronto, ON, Canada; Dunlap Institute of Astronomy \& Astrophysics, University of Toronto, Toronto, ON, Canada}\\
  Department of Statistical Sciences\\
  University of Toronto \\
  Toronto, ON, Canada\\
  \texttt{j.speagle@utoronto.ca}
   \And
   Song Huang (黄崧)\\
   Department of Astronomy\\
   Tsinghua University \\
   Beijing, China 100084 \\
   \texttt{shuang@tsinghua.edu.cn} \\
   \And
   Yuan-Sen Ting (丁源森)\thanks{Secondary Affiliations: School of Computing, Australian National University, Acton, ACT 2601, Australia; Department of Astronomy, The Ohio State University, Columbus, USA}\\
   Research School of Astronomy \& Astrophysics\\
   Australian National University\\
   Cotter Rd., Weston, ACT 2611, Australia\\
   \texttt{yuan-sen.ting@anu.edu.au}\\
   \And
   Zheng Cai (蔡峥)\\
   Department of Astronomy \\
   Tsinghua University \\
   Beijing, China 100084 \\
   \texttt{zcai@mail.tsinghua.edu.cn}
}

\begin{document}
\begin{CJK*}{UTF8}{gkai}

\maketitle

\begin{abstract}
  We present \NAME, a novel method that integrates cutting-edge normalizing flow techniques into a mixture density estimation framework, enabling the effective use of heterogeneous training data for photometric redshift inference. Compared to previous methods, \NAME demonstrates enhanced robustness for both point estimation and distribution reconstruction by leveraging normalizing flows for density estimation and incorporating careful uncertainty quantification. Moreover, \NAME offers unique interpretability by explicitly disentangling contributions from multi-source training data, which can facilitate future weak lensing analysis by providing an additional quality assessment. As probabilistic generative deep learning techniques gain increasing prominence in astronomy, \NAME should become an inspiration for handling heterogeneous training data while remaining interpretable and robustly accounting for observational uncertainties.
\end{abstract}

\section{Introduction}\label{sec:intro}
%\vspace{-0.1in}
Redshift measures cosmic distances and Universe expansion, making it fundamental in astrophysics \citep{DODELSON2003, WEINBERG2008, MO2010}. Cosmology \citep{ALAM2021, DALAL2023, SUGIYAMA2023} and extragalactic science \citep{HELTON2023} both count on accurate redshifts with a wide range of sensitivity requirement. For example, weak lensing studies, which probe cosmological structure growth and expansion history by tracing dark matter cosmic webs through the distortion of galaxy shapes \citep{RACHEL2018}, are currently limited by redshift errors \citep{ABRUZZO2019, YUAN2019, ALEX2022}.

High-precision redshifts require expensive, biased spectroscopic observations (spec-$z$/grism-$z$/prism-$z$). In contrast, photometric redshifts (photo-$z$) cover wider luminosity ranges but lack spectroscopic precision. In ongoing and future surveys like LSST \citep{IVEZIC2019}, most galaxies will have only photometric data. Integrating spectroscopic and photometric data is therefore crucial for photo-$z$ inference. 

% current available methods
Photo-$z$ estimation techniques fall into two categories: template-fitting \citep{BENITEZ2000, BRAMMER2008, LAUR2022}, which matches photometry to spectroscopic or physical model templates, and machine learning \citep{SADEH2016, BOEIS2017, JONES2022}, which trains supervised models to map photometry to reference redshifts. To maximize information gain for science study, photo-$z$ estimation requires synthesizing all available data to achieve wide redshift coverage, high precision, and minimal selection bias. 

% heterogeneous dataset problem 
Stitching high-quality spectroscopic redshifts \citep{AHUMADA2020}, medium-quality grism/prism redshifts \citep{BRAMMER2012, COOL2013}, and lower-quality photometric redshifts \citep{LAIGLE2016} enables full exploitation of deep galaxy surveys like HSC-SSP \citep{AIHARA2018} for photo-$z$ estimation. To help accomplish this, we propose \NAME -- an integrative framework that stitches heterogeneous training samples for photo-$z$ inference using normalizing flows \citep{KOBYZEV2019}. As a generalized extension of \texttt{frankenz}\citep{SPEAGLE2019} for future large-scale sky surveys, \NAME: (1) improves photo-$z$ inference, refines both point estimates and redshift probability density estimation; (2) interprets heterogeneous datasets and exerts uncertainty control; (3) more efficiently scales to high-dimensional feature spaces and large datasets.

%\vspace{-0.1in}
\section{Method}\label{sec:method}
%\vspace{-0.05in}

\begin{figure*}[t]
    \centering
    \includegraphics[scale=0.5]{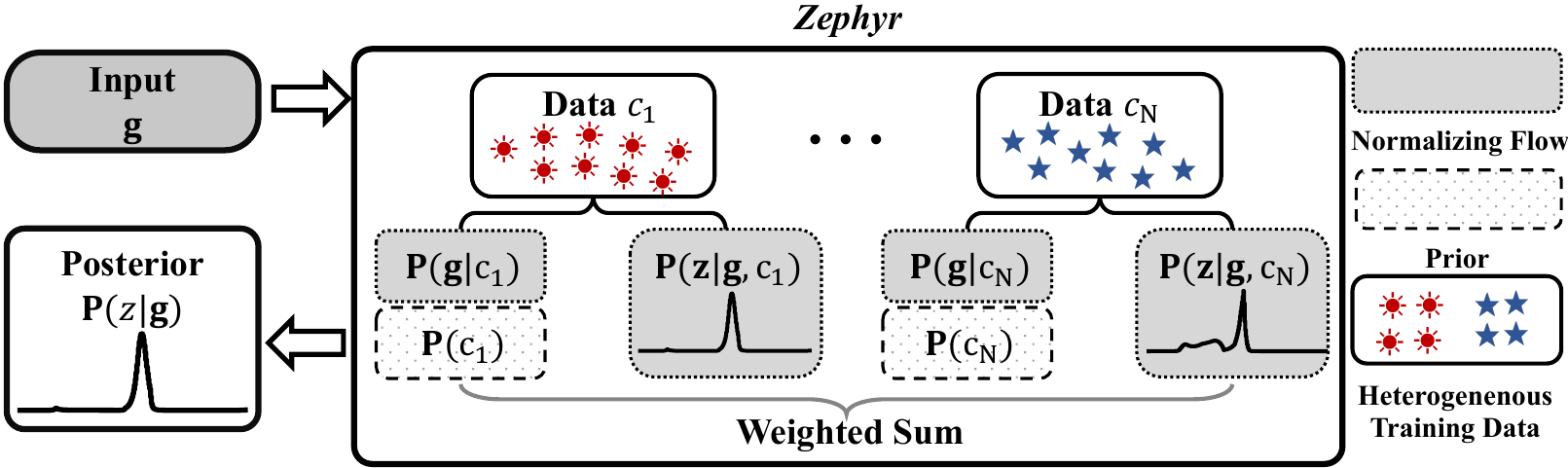}
    \caption{Model architecture. \NAME integrates normalizing flows and mixture density estimation to give both accurate and interpretable photo-$z$ inference with noisy, heterogeneous training data.}
    \label{fig:schematic}
\end{figure*}

%\vspace{-0.1in}
%\subsection{Statistical Framework}
%\vspace{-0.05in}

Our \NAME framework combines heterogeneous training data for photo-$z$ inference via mixture density estimation as shown in Figure~\ref{fig:schematic}. The photo-$z$ posterior probability density function (PDF) $\mathrm{P}(z|\mathbf{g}, \boldsymbol{\sigma})$ for redshift $z$ given photometry $\mathbf{g}$ and uncertainty $\boldsymbol{\sigma}$ is expressed as a weighted sum of posteriors from distinct categories $\mathrm{P}(z|\mathbf{g}, c_i)$, with each category $c_i$, $i=1,2,\dots,\mathrm{M}$ weighted by $\mathrm{P}(\mathbf{g}|c_i)\mathrm{P}(c_i)$ as shown in Equation~\ref{eq:framework}. 
%\vspace{-0.05in}
\begin{equation}
    \mathrm{P}(z|\mathbf{g}, \boldsymbol{\sigma}) = \sum_{i=1}^{\mathrm{N}} \mathrm{P}(z|\mathbf{g}, \boldsymbol{\sigma}, c_i) \mathrm{P}(c_i|\mathbf{g}, \boldsymbol{\sigma})
    \propto \sum_{i=1}^{\mathrm{N}} \mathrm{P}(z|\mathbf{g}, \boldsymbol{\sigma}, c_i) \mathrm{P}(\mathbf{g}|\boldsymbol{\sigma}, c_i) \mathrm{P}(c_i)
\label{eq:framework}
\end{equation}
We treat the category prior weights $\mathrm{P}(c_i)$ as latent variables in our mixture density model. $\mathrm{P}(z|\mathbf{g}, \boldsymbol{\sigma}, c_i)$ and $\mathrm{P}(\mathbf{g}|\boldsymbol{\sigma}, c_i)$ are estimated using normalizing flows, which are designed to transform simple distributions into complex ones via invertible, differentiable functions, enabling efficient sampling and density estimation (\citep{DURKAN2019}). \NAME's use of normalizing flows over nearest neighbor methods for density estimation improves performance and high-dimensional scaling over \texttt{frankenz}.

Denoting $\mathbf{g}^*$ as the true underlying photometric data, $\mathrm{P}(z|\mathbf{g}, \boldsymbol{\sigma}, c_i)$ can be formulated as: 
\begin{equation}
\mathrm{P}(z|\mathbf{g}, \boldsymbol{\sigma}, c_i) = \int\mathrm{P}(z|\mathbf{g}^*, c_i)\mathrm{P}(\mathbf{g}^*|\mathbf{g}, \boldsymbol{\sigma})\,\mathrm{d}\Omega_\mathbf{g}^*
= \int\mathrm{P}(z|\mathbf{g}^*, c_i)\mathrm{P}(\mathbf{g}|\mathbf{g}^*, \boldsymbol{\sigma}) \frac{\mathrm{P}(\mathbf{g}^*)}{\mathrm{P}(\mathbf{g})}\,\mathrm{d}\Omega_\mathbf{g}^*
\label{eq:uncertainty:1} 
\end{equation}
where $\Omega_\mathbf{g}^*$ denotes the photometric space and $\mathrm{P}(\mathbf{g}|\mathbf{g}^*, \boldsymbol{\sigma})$ follows $\mathcal{N}(\mathbf{g}^* ; \boldsymbol{\sigma})$. Although the unknown prior distribution over the true photometry $\mathrm{P}(\mathbf{g}^*)$ may be complex, in this work we approximate it as uniform. We justify this approximation for two reasons: (1) for high signal-to-noise cases, $\mathrm{P}(\mathbf{g}|\mathbf{g}^*,\boldsymbol{\sigma})$ is concentrated around $\mathbf{g}^*$ and so $\mathrm{P}(\mathbf{g}^*)$ would be roughly constant; (2) for low signal-to-noise cases, a uniform $\mathrm{P}(\mathbf{g}^*)$ would generally lead to slightly broader PDFs. 

Approximating Equation~\ref{eq:uncertainty:1} through Monte Carlo integration \citep{LEJA2020, LEJA2022, WANG2022, WANG2023} with $\mathrm{M}$ samples drawn from this distribution, we have:
\begin{equation}
        \mathrm{P}(z|\mathbf{g},\boldsymbol{\sigma}, c_i) \approx \frac{1}{\mathrm{M}} \sum_{j=1}^{\mathrm{M}} \mathrm{P}(z|\mathbf{g}^*, c_i)
\end{equation}
Similarly, $\mathrm{P}(\mathbf{g}|\boldsymbol{\sigma}, c_i)$ can be approximated the same way to get: 
\begin{equation} 
\mathrm{P}(\mathbf{g}|\boldsymbol{\sigma}, c_i) = \int \mathrm{P}(\mathbf{g}|\mathbf{g}^*, \boldsymbol{\sigma}) \mathrm{P}(\mathbf{g}^*|c_i)\,\mathrm{d}\Omega_\mathbf{g}^*  \approx \frac{1}{\mathrm{M}}\sum_{j=1}^\mathrm{M} \mathrm{P}(\mathbf{g}^*_j|c_i), 
\label{eq:uncertainty:2}
\end{equation}
where $\mathbf{g}^*_j,\, j=1,2,3,\dots,\mathrm{M}$ follow the same distribution as in Equation~\ref{eq:uncertainty:1}. This approach allows us to learn the intrinsic distributions for $\mathrm{P}(z|\mathbf{g}, \boldsymbol{\sigma},c_i)$ and $\mathrm{P}(\mathbf{g}|\boldsymbol{\sigma}, c_i)$, providing a robust method to handle uncertainty. Equation~\ref{eq:uncertainty:1} and~\ref{eq:uncertainty:2} facilitate finer uncertainty control with the normalizing flow and are used during both model training and inference.

\begin{figure}[t]
\centering
\subfigure[Circle\label{subfig:circle}]{\includegraphics[width=0.4\linewidth]{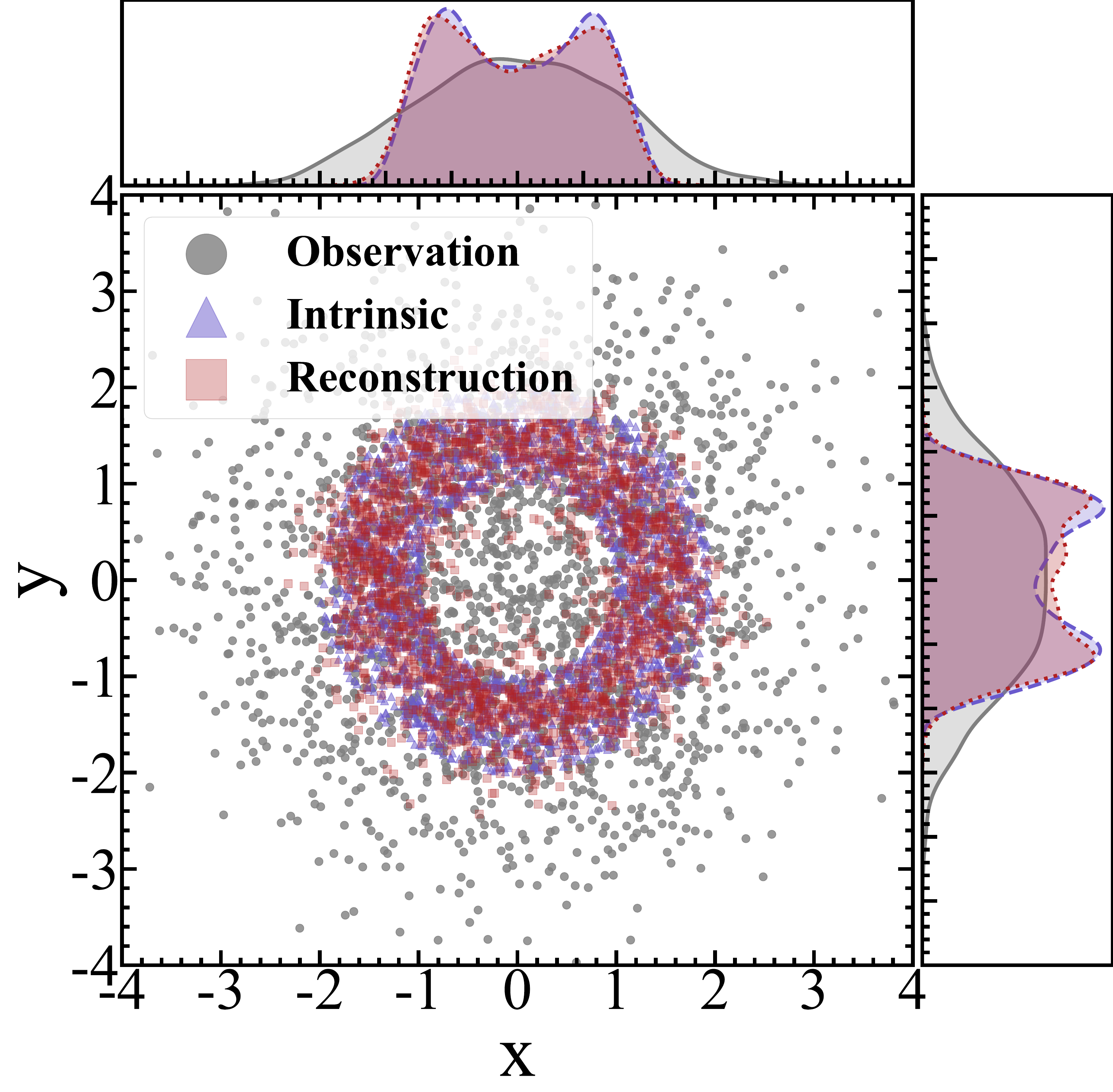}}
\subfigure[Moon\label{subfig:moon}]{\includegraphics[width=0.4\linewidth]{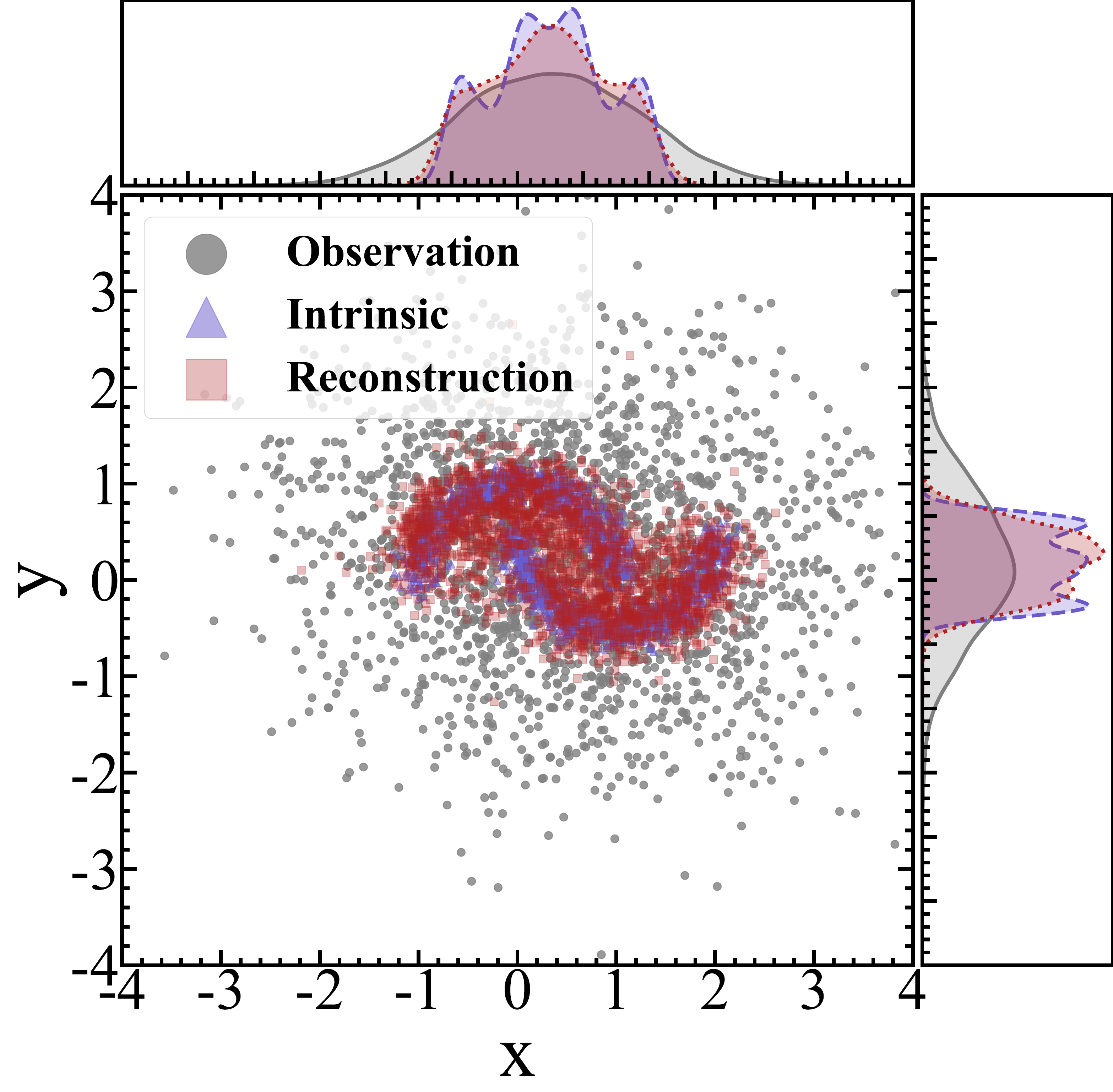}}
\caption{Recovering distributions from noisy data. The proposed method effectively estimates densities on circle (left) and moon (right) datasets despite low signal-to-noise ratios. \label{fig:uncer}}
\end{figure}

We showcase the quality of our uniform $\mathrm{P}(\mathbf{g}^*)$ assumption for low signal-to-noise cases on two toy datasets. In Figure~\ref{fig:uncer}, the intrinsic distributions (blue triangles) are a circle \ref{subfig:circle} and a double moon \ref{subfig:moon}. Gaussian noise $\mathcal{N}(0, 1)$ is added to simulate low signal-to-noise observations (grey dots). The experiment shows that while the uniform prior may smear the true distribution slightly, it remains highly effective for recovering the underlying true density (red squares).

%\vspace{-0.3in}
\section{Data}\label{sec:data}
%\vspace{-0.1in}

We analyze a dataset with two reference redshifts: a collection of high-quality spec-$z$'s/grism-$z$'s/prism-$z$'s (high-confidence, biased to brighter objects) and lower-quality photo-$z$'s (lower-confidence, fainter objects). The photometry is taken from HSC-SSP survey PDR3 \citep{AIHARA2022} in the {\it grizy} filters. We cross-match HSC PDR3 against other surveys (see Table~\ref{tab:data}) to obtain our collection of reference redshifts. COSMOS2015 \citep{LAIGLE2016} provides 30-band photo-$z$. Our final dataset includes $129,449$ photo-$z$ sources from COSMOS2015 and $21,591$ spec-$z$ sources from spectroscopic surveys. We split the data 90/5/5 percent portions for training/validation/testing. 

\begin{table}[htbp!]
\begin{center}
\begin{small}
\begin{tabular}{lccc}
\toprule
Method & Bias ($b$) & Scatter ($\sigma$) & Outlier Rate ($\eta$)\\
\midrule
\NAME & -0.003 & \textbf{0.053} & \textbf{0.198}\\
\texttt{frankenz} & \textbf{0.001} & 0.080 & 0.264\\ 
\texttt{EAzY} & -0.011 & 0.170 & 0.458\\
\bottomrule
\end{tabular}
\end{small}
\end{center}
\caption{Comparison on performance over the test data. The normalizing flow-based \NAME model achieves a much lower scatter and outlier rate compared to \texttt{frankenz} and \texttt{EAzY}.}
\label{tab:result}
%\vskip -0.3in
\end{table} 

\section{Experiment and Result}\label{sec:result}
%\vspace{-0.1in}
\subsection{Experiment Settings}

We use neural spline flows \citep{DURKAN2019} for density estimation. We set the spec-$z$ and photo-$z$ prior probabilities to their relative portions in the training data. We investigate: (a) \NAME's photo-$z$ inference capability; and (b) its interpretability in quantifying contributions from different training samples. We compare to (1) \texttt{frankenz}\footnote{\href{https://github.com/joshspeagle/frankenz}{https://github.com/joshspeagle/frankenz}} \citep{SPEAGLE2019}, a Bayesian nearest neighbor method conceptually similar to \NAME, and (2) \texttt{EAzY}\footnote{\href{https://github.com/gbrammer/eazy-photoz}{https://github.com/gbrammer/eazy-photoz}} \citep{BRAMMER2008}, a widely-used template-fitting method, with the \texttt{sfhz}\footnote{\href{https://github.com/gbrammer/eazy-photoz/tree/master/templates/sfhz}{https://github.com/gbrammer/eazy-photoz/tree/master/templates/sfhz}} template.

%\texttt{frankenz} {https://github.com/joshspeagle/frankenz}}, a Bayesian nearest neighbor method, combines heterogeneous training by propagating uncertainties. It excels at reconstructing redshift distributions and quantifying training contributions critical for cosmology. However, its simplicity limits the quality of its inference. In addition, nearest neighbor search can be inefficient and can struggle with less-representative training contributions. 

%\subsection{Verify Uncertainty Quantification on Toy Dataset}\label{subsec:toy}

\subsection{Assessing the Quality of Photo-$z$ Inference}

To assess photo-$z$ quality, we define the scaled residual $\Delta z = (z_{pred}-z_{ref})/(1+z_{ref})$, bias $b = \mathrm{Median}(\Delta z)$, scatter $\sigma=1.4826\times\mathrm{Median}(|\Delta z - b|)$, and outlier rate $\eta$ (outliers have $|\Delta z|>0.15$) \citep{EUCLID2020,NISHIZAWA2020}. As shown in Table~\ref{tab:result}, \NAME exhibits lower scatter and outlier rate than \texttt{frankenz} and \texttt{EAzY}, attributable to the high expressiveness of normalizing flows used. For distribution reconstruction, \NAME shows comparable probability integral transform (PIT) performance to \texttt{frankenz} and outperforms \texttt{EAzY} (Figure~\ref{fig:pop}), demonstrating viability for cosmology \citep{ABRUZZO2019,SCHMIDT2020}.

\subsection{Interpreting Heterogeneous Datasets}

\begin{figure}
\centering
\subfigure[Population-level Inference\label{fig:pop}]{\includegraphics[width=0.48\linewidth]{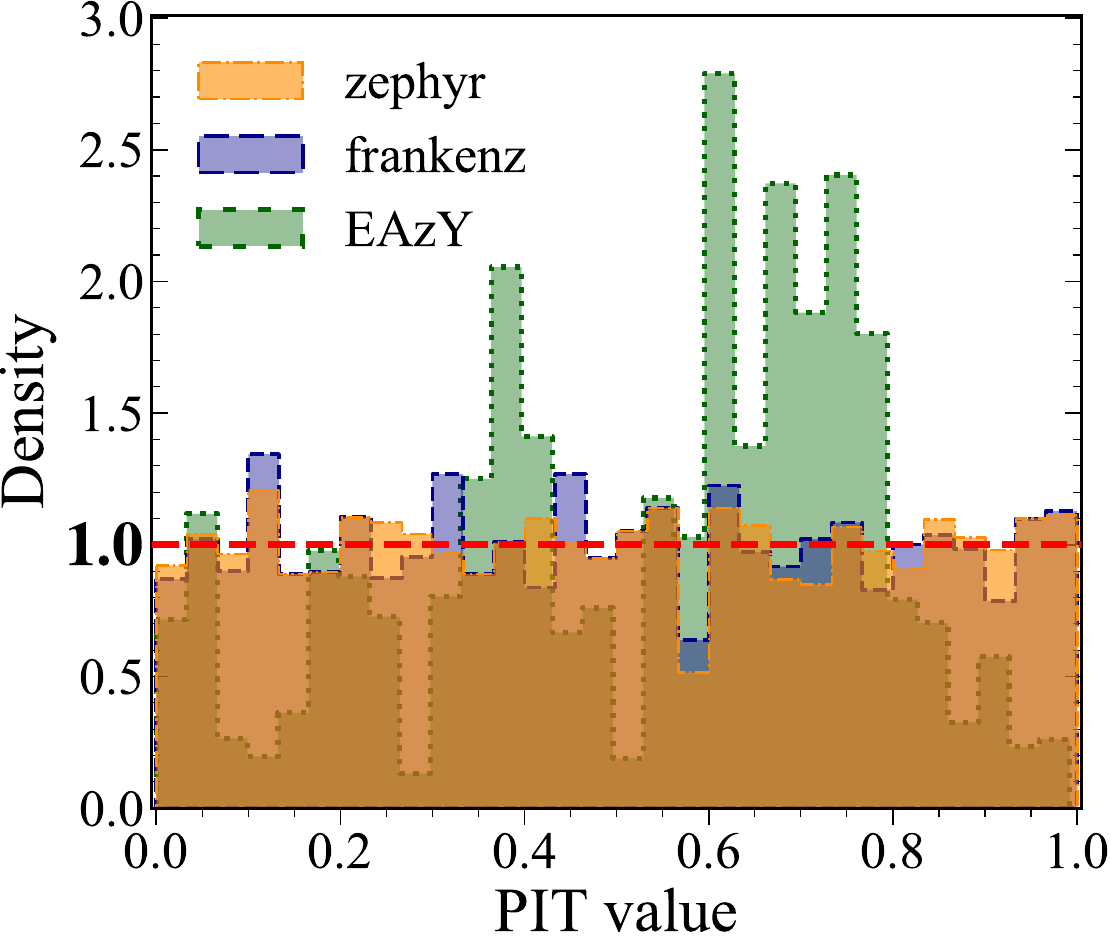}}
\subfigure[Interpreting Heterogeneous Training Data\label{fig:interp}]{\includegraphics[width=0.48\linewidth]{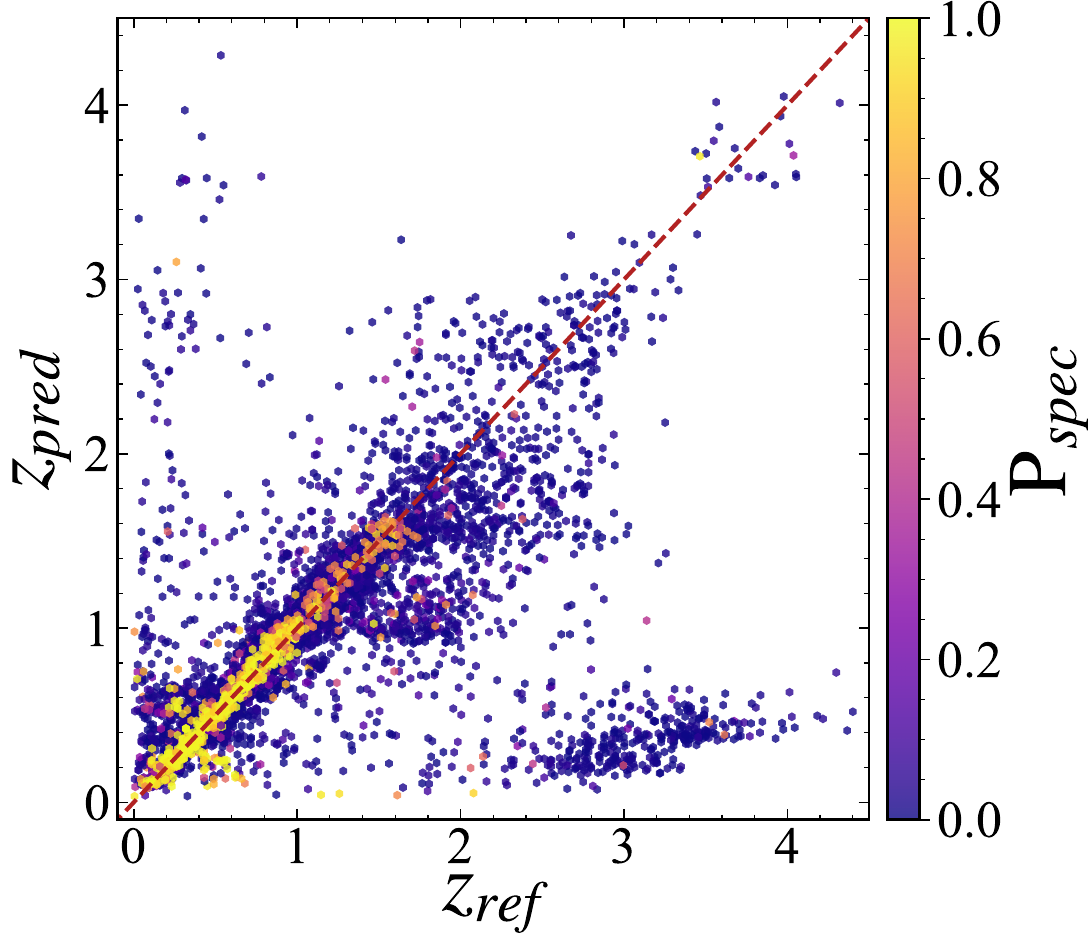}}
\caption{(a) PIT distributions for different models, demonstrating \NAME's  efficacy in reconstructing the redshift distribution as its PIT value closely follows a uniform distribution $\mathcal{U}(0, 1)$; (b) \NAME point estimates versus reference redshifts color-coded by spec-$z$ training sample contributions. \NAME's results clearly show that the high-redshift end of the distribution and most redshift outliers predominantly are made up of predictions with little spec-$z$ contributions.}
\end{figure}

Assessing impacts of heterogeneous data is crucial in cosmology. We show the \NAME model has strong interpretability, facilitating disambiguation of contributions from disparate datasets. We define $\mathrm{P}_{spec}$ as the proportion of spec-$z$ samples contributing to photo-$z$ estimation:
\begin{equation}
\mathrm{P}_{\rm spec} = \frac{\mathrm{P}(\mathbf{g}|c_1)\mathrm{P}(c_1)}{\mathrm{P}(\mathbf{g}|c_1)\mathrm{P}(c_1) + \mathrm{P}(\mathbf{g}|c_2)\mathrm{P}(c_2)}. 
\end{equation}
Figure~\ref{fig:interp} shows the spec-$z$ training samples contribute less to high-redshift sources due to limited depth of spectroscopic surveys. In addition, we see predictions dominated by photo-$z$ training samples also have larger scatter and outlier rates due to imprecise reference redshifts and noisier input photometry. \NAME exhibits superior interpretability in disentangling contributions from heterogeneous datasets, enabling unique quality assessment opportunities for cosmology analysis. 

\section{Broader Impact}\label{sec:impact}
%\vspace{-0.1in}
As we enter an era of ultra-precise cosmology, accurate photo-$z$ inference is critical. To achieve this, we need advanced algorithms to maximize photo-$z$ precision and understand training data impacts. \NAME enables both - its flexible framework handles high-dimensional data while maintaining interpretability. Moreover, normalizing flows enable data-driven inferences for astrophysics. We will expand \NAME into a versatile framework for broader applications like stellar mass inference. Such capabilities will prove valuable for upcoming surveys.

\section*{Acknowledgement}\label{sec:acknowledge}
The authors thank Diana Blanco, Alexie Leauthaud, and Yifei Luo (罗逸飞) from the Department of Astronomy and Astrophysics at the University of California, Santa Cruz for fruitful discussions regarding potential applications of \texttt{zephyr} in weak lensing analysis. The authors also thank the anonymous reviewers for their valuable comments.

\bibliography{main}
\bibliographystyle{neurips_2021}

\appendix

\section{Appendix - Dataset Composition}
We detail the data utilized in this work here. 

\begin{table}[h]
\label{tab:data}
\centering
\begin{small}
\begin{tabular}{llc}
\toprule
Name & Category & Reference\\
\midrule
COSMOS2015 & photo-$z$ & \citep{LAIGLE2016}\\
PRIMUS & prism-$z$ & \citep{COOL2013}\\
3D-HST  & grism-$z$ & \citep{BRAMMER2012} \\
SDSS DR16 & spec-$z$ & \citep{AHUMADA2020} \\
C3R2 & spec-$z$ & \citep{MASTERS2017, MASTERS2019} \\
DEIMOS & spec-$z$ & \citep{HASINGER2018} \\
MOSDEF & spec-$z$ & \citep{KRIEK2015}\\
LEGA-C & spec-$z$ & \citep{ARJEN2021} \\
\bottomrule
\end{tabular}
\end{small}
\vskip 0.1in
\caption{Reference redshift catalogs in this work. COSMOS2015 provides unique photo-$z$ samples from 30-band photometry.}
\end{table}

\end{CJK*}
\end{document}